\documentclass{article}
\usepackage{spconf,amsmath,graphicx,color,url}

\def\L{{\cal L}}
\title{VisualTTS: TTS with Accurate lip-speech synchronization \\ for automatic voice over}
\name{Junchen Lu \textsuperscript{1}, Berrak Sisman \textsuperscript{2}, Rui Liu \textsuperscript{1,2}, Mingyang Zhang \textsuperscript{1}, Haizhou Li \textsuperscript{1,3}
\thanks{Speech samples: \url{https://ranacm.github.io/VisualTTS-Samples/}}
\thanks{This work is supported by A*STAR under its RIE2020 Advanced Manufacturing and Engineering Domain (AME) Programmatic Grant (Grant No. A1687b0033, Project Title: Spiking Neural Networks).}
}
\address{\textsuperscript{1} Department of Electrical and Computer Engineering, National University of Singapore \and
\textsuperscript{2} Information Systems Technology and Design, Singapore University of Technology and Design \and
\textsuperscript{3} The Chinese University of Hong Kong, Shenzhen, China}
\begin{document}
%
\maketitle
\begin{abstract}
In this paper, we formulate a novel task to synthesize speech in sync with a silent pre-recorded video, denoted as automatic voice over (AVO). Unlike traditional speech synthesis, AVO seeks to generate not only human-sounding speech, but also perfect lip-speech synchronization. A natural solution to AVO is to condition the speech rendering on the temporal progression of lip sequence in the video. 
We propose a novel text-to-speech model that is conditioned on visual input, named \textit{VisualTTS}, for accurate lip-speech synchronization. The proposed VisualTTS adopts two novel mechanisms that are 1) textual-visual attention, and 2) visual fusion strategy during acoustic decoding, which both contribute to forming accurate alignment between the input text content and lip motion in input lip sequence. Experimental results show that VisualTTS achieves accurate lip-speech synchronization and outperforms all baseline systems.

\end{abstract}
\section{Introduction}
\label{sec:intro}

Automatic voice over (AVO) aims to deliver speech that voice-synchronizes with a silent pre-recorded video.
An AVO system takes a silent video of a spoken utterance and its text script as the input, and generate natural speech that synchronizes with lip motion, emotional states, and dialogue scenarios in the video automatically.
AVO technology will transform the way the movie industry conducts voice over. It will also enable new applications in entertainment, education, and business.

Text-to-speech synthesis (TTS) is the task of synthesizing speech from text input. With the advent of deep learning,
end-to-end neural TTS systems are able to produce high-quality synthetic speech.
In these techniques, the key idea is to integrate the conventional TTS pipeline into a unified encoder-decoder network and to learn the mapping in the $<$text, wav$>$ pair \cite{liu2021expressive}. 
Successful implementations include Tacotron 1/2 \cite{wang2017tacotron, shen2018natural}, Transformer TTS \cite{li2019neural}, FastSpeech 1/2 \cite{ren2019fastspeech, ren2020fastspeech} and their variants \cite{liu2021expressive, yasuda2019investigation, cooper2020zero}.
Together with neural vocoders \cite{oord2016wavenet, kalchbrenner2018efficient}, they can generate impressive natural-sounding speech.

Motivated by the study of neural TTS, a natural solution to AVO is to build a TTS system by taking text script as input, and conditioning on the temporal progression of lip movement and facial expression. One of the challenges is that humans are sensitive to  audio-video mismatch. A minor mismatch may seriously affect the perceived speech quality, and intelligibility. A general purpose TTS doesn't guarantee such lip-speech synchronization as no visual information is taken into consideration.

Audio-video synchronization has been exploited in multi-modal signal processing, such as multi-modal speech recognition \cite{afouras2018deep}, and multi-modal speech separation \cite{ephrat2018looking}.
For example, Afouras et al. \cite{afouras2018deep} studied the use of Transformer \cite{vaswani2017attention} 
for audio-visual information fusion, which achieves remarkable performance in multi-modal speech recognition. Pan et al. \cite{pan2021muse} proposed a multi-modal speaker extraction network to introduce lip-speech synchronization cues obtained from lip image sequence as the reference signal for speech extraction from a target speaker. 

In this paper, we propose a TTS framework leveraging visual information (VisualTTS) with textual-visual attention and visual fusion strategy, which can learn the accurate alignment between the text script and the lip motion in input lip image sequence obtained from a video clip of spoken utterance. We conduct experiments on GRID dataset \cite{cooke2006audio}. VisualTTS achieves accurate lip-speech synchronization and outperforms all baseline systems.

The main contributions of this paper include: 1) we formulate the AVO research problem and propose a novel neural model to incorporate visual information into TTS; 2) we propose two novel mechanisms, textual-visual attention and visual fusion strategy, to achieve accurate lip-speech synchronization. To our best knowledge, this is the first in-depth study of automatic voice over in speech synthesis. 


The rest of the paper is organized as follows: Section 2 presents the related work of this paper; Section 3 elaborates the model architectures; Section 4 describes details of our experiments; Section 5 provides conclusion of this paper.

\section{Related work}
\label{sec:related}

\subsection{Multi-modal speech synthesis}

There have been studies on speech synthesis with multi-modal information
such as image-to-speech \cite{hsu2020text, effendi2021end}, video-to-speech \cite{prajwal2020learning, mira2021end} and automatic dubbing \cite{federico2020evaluating}.  The AVO task is a completely new multi-modal speech synthesis task, which has not been investigated in depth. AVO takes a text script and a silent video clip as input to generate a speech audio that synchronizes with the lip motion and facial expression in video.  

\begin{figure}
    \centering
    \includegraphics[scale=0.6]{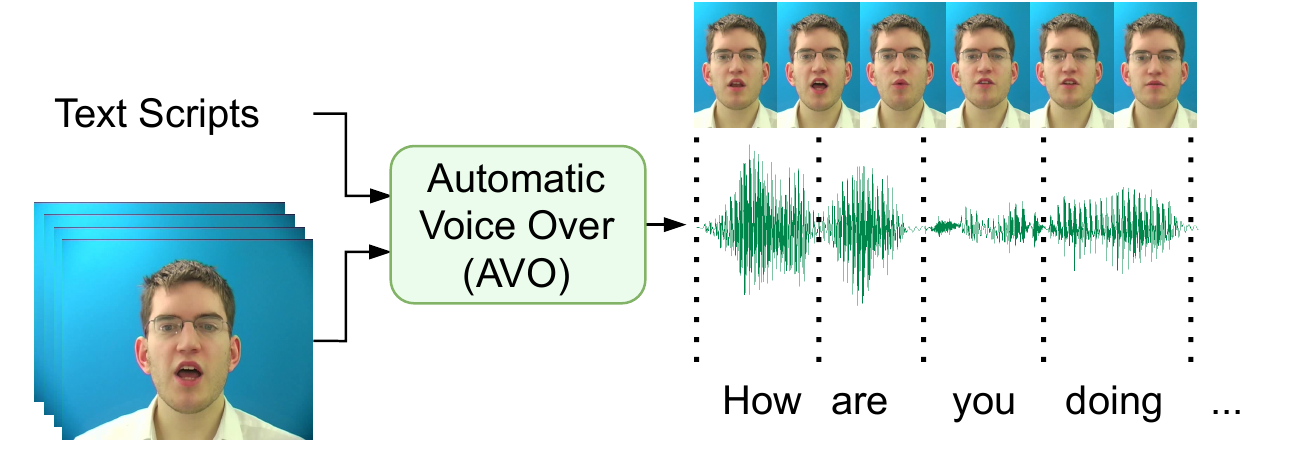}
    \vspace{-4mm}
    \caption{The typical workflow of automatic voice over: An AVO framework takes video and text script as input, and generates speech audio in sync with video.}
    \vspace{-4mm}
    \label{fig:AVO}
\end{figure}

An AVO workflow is illustrated in Fig. \ref{fig:AVO}. It differs from other multi-modal speech synthesis tasks in many ways. To start with, image-to-speech \cite{hsu2020text, effendi2021end} seeks to generate caption speech from an image, while video-to-speech \cite{prajwal2020learning, mira2021end} aims to reconstruct speech signal from silent video of utterances spoken by people. Both tasks take visual information as the sole input to predict the speech output, while AVO receives both text and video as input. The study on automatic dubbing \cite{federico2020evaluating} is essentially to generate speech in one language for a video in another language, where machine translation plays a keep role while lip-speech synchronization is not the main focus. 


In an AVO task, visual information learning and representation are required to synchronize the synthesized voice with the video input,  which will be the focus of this paper.

\vspace{-2mm}
\subsection{Visual embedding}
\label{sec:visual}
Video clips contain important information that can be useful for speech synthesis such as lip motion, facial expression and emotional states \cite{pan2020multi, prajwal2020learning}.
Appropriate rendering of phonetic duration in output speech depends on accurate lip-speech synchronization. As the modeling of lip-speech synchronization is built on the characterization of lip motion and speech signals~\cite{pan2021muse, chen2021correlating}, feature representation of lip motion from video is critically important.

Visual embedding has been successfully used in speech research. For lip-reading task, which is also known as visual speech recognition, the use of visual embedding has shown to provide useful information by condensing the lip motion information in the video \cite{afouras2018deep, assael2016lipnet}. Another example is the audio-visual speech enhancement task, in which Chen et al. \cite{chen2021correlating} proposed to fuse visual embedding extracted in a lip-reading task with audio embedding to provide lip-speech correlation information.



Inspired by the recent success in visual embedding, we propose to use visual embedding extracted by a lip-reading network to guide the duration alignment in our VisualTTS for accurate lip-speech synchronization.



\section{VisualTTS}

\label{sec:model}


We formulate the AVO problem and propose a visual TTS solution next. With the motivation of generating speech in accurate synchronization with video, in VisualTTS, we propose a novel textual-visual attention and a visual fusion strategy for leveraging lip-speech synchronization information obtained from lip image sequence.

\subsection{Overall architecture}
As shown in Fig.~\ref{fig:modelarchitecture}, the overall architecture of VisualTTS consists of visual encoder, textual encoder, speaker encoder, visual-guided aligner, acoustic decoder and WaveNet vocoder~\cite{oord2016wavenet}. 

The visual encoder aims to learn the visual embedding $\alpha$ to represent the lip motion information of the given lip image sequence.
The textual encoder takes the input script as input to generate the textual embedding $\beta$ .
The speaker encoder seeks to encode the speaker ID into an utterance level speaker embedding $\gamma$ .
Textual embedding and visual embedding are then sent to visual-guided aligner for textual-visual alignment learning.
The outputs of visual-guided aligner are decoded by the acoustic decoder into mel-spectrogram features, which are then converted to audio waveform using a pre-trained WaveNet vocoder~\cite{oord2016wavenet, wang2018comparison}.

The textual encoder consists of a character embedding layer and a CBHG-LSTM module, which is similar to that of Tacotron \cite{wang2017tacotron}.
We will introduce the visual encoder, speaker encoder, visual-guided aligner and acoustic decoder in detail next.

\vspace{-2mm}

\subsubsection{Visual encoder}

The AVO task takes text and video as input, hence as a pre-processing step, we obtain lip image sequence $\mathcal{L}$ by cropping the lip region from frames of video. We note that each lip image corresponds to one frame of video. We then propose to use a visual encoder to exploit the visual cues from lip image sequence, as shown in the left panel of Fig. \ref{fig:modelarchitecture}.


The visual encoder consists of a 3D convolutional (Conv3D) layer and a ResNet-18 block \cite{wu2019time}. Such an architecture has shown to be effective in the lip-reading task to learn the lip motion information in the video \cite{afouras2018conversation}. 
The visual encoder takes $\mathcal{L}$ as input and outputs the visual embedding $\alpha$ for each frame of lip image sequence $\mathcal{L}$.


We note that all modules of visual encoder are pre-trained in a lip-reading task, in a similar way that is reported in \cite{wu2019time}. In other words, during VisualTTS training, all weights of the visual encoder are fixed.

\subsubsection{Speaker encoder}
\label{sec:spk}
\begin{figure}[!t]
    \centering
    \includegraphics[scale=0.78]{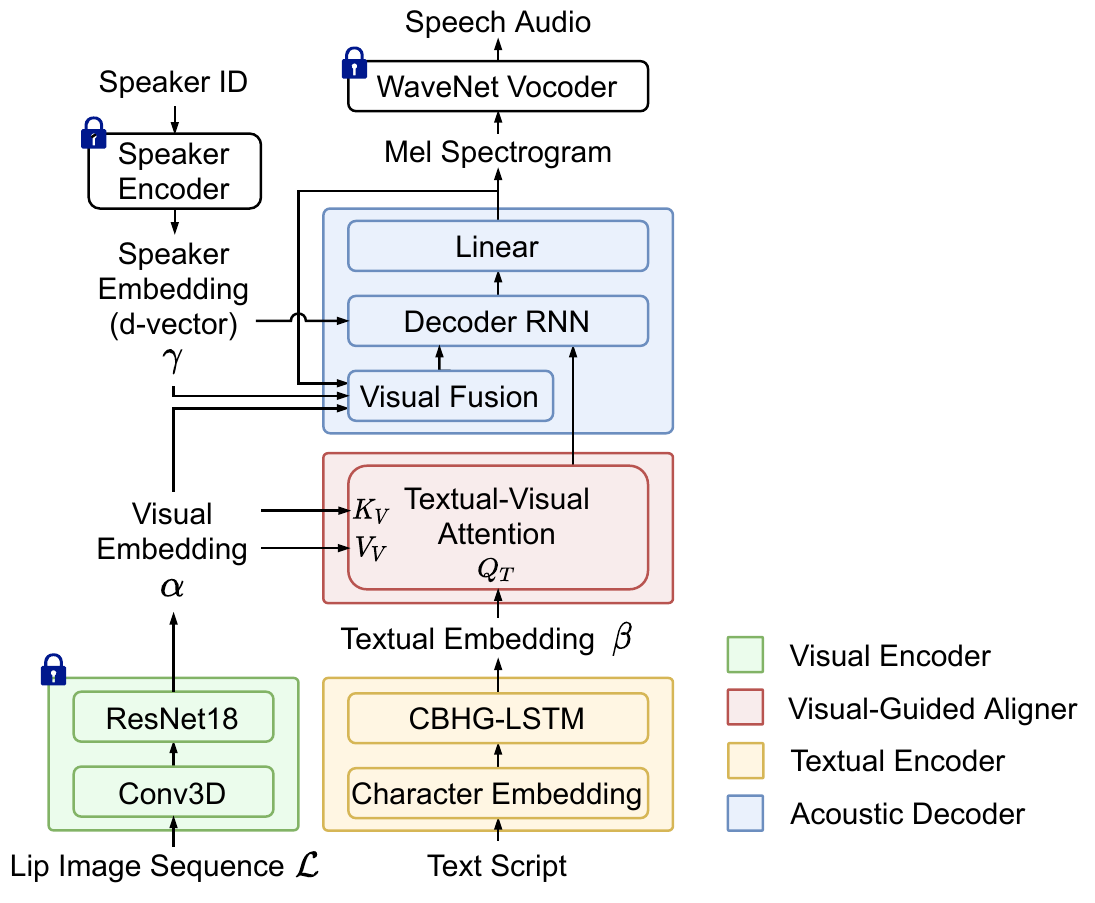}
    \vspace{-5mm}
    \caption{Model architecture of the proposed VisualTTS, that consists of a visual encoder, a textual encoder, a visual-guided aligner and an acoustic decoder. Pre-trained blocks are denoted with a lock. }
    \label{fig:modelarchitecture}
\end{figure}

VisualTTS aims to achieve multi-speaker speech synthesis, hence we use a speaker encoder as shown in Fig. \ref{fig:modelarchitecture} to obtain the speaker embedding for a given speaker ID, which is a unique integer assigned to each speaker.

We note that the speaker encoder adopts a lookup table to match d-vector $\gamma$ obtained by a pre-trained speaker verification model \cite{variani2014deep}.

\subsubsection{Visual-guided aligner}

The visual-guided aligner consists of a textual-visual attention (TVA) mechanism to align cross-modal information, namely textual and visual information.


Specifically, the output of visual encoder, visual embedding $\alpha$, is passed to TVA as key $K_V$ and value $V_V$. Textual embedding $\beta$ is passed to TVA as query $Q_T$. 
A multi-head scaled dot-product attention \cite{vaswani2017attention} is used for our implementation of TVA.
The textual-visual context is given by:

\vspace{-7mm}
\begin{subequations}
\begin{align}
    C(Q_T,K_V,V_V)=&softmax(\frac{Q_TK_V^T}{\sqrt{d_{K_V}}})V_V \\= & softmax(\frac{\beta \alpha^T}{\sqrt{d_{\alpha}}})\alpha
\end{align}
\end{subequations}
where 
$d_{K_V}$ is the dimension of $\alpha$.


Since the content of speech is determined solely by its corresponding text script, 
speech can be synchronized with lip motion accurately if the content of speech matches with lip motion information.
In such a way, TVA captures long-term information for textual-visual dependency and learns the alignment between textual embedding and visual embedding, 
thus helps to yield speech well aligned with lip motion.





\subsubsection{Acoustic decoder}

The acoustic decoder consists of a visual fusion layer, and the decoder from Tacotron \cite{wang2017tacotron} that consists of an attention-based recurrent neural network (RNN), and a linear layer.

In practice, the length of mel-spectrogram is of certain ratio to the length of visual embedding, since speech audio and video are temporally synchronized. 
Each frame of the mel-spectrogram can be indexed to its corresponding video frame according to this ratio.
In each time step of acoustic decoding, a frame of mel-spectrogram feature is concatenated with its corresponding visual embedding by the visual fusion layer.
The purpose is to leverage the temporal correlation between visual embedding and mel-spectrogram.
The concatenated representation is added with the speaker embedding to form a multi-modal representation~\cite{cooper2020zero}, which is then projected to a multi-modal hidden sequence as output of the visual fusion layer. During acoustic decoding, the output of TVA is concatenated with speaker embedding \cite{cooper2020zero} and passed to the rest part of the decoder along with the visual fusion output, and then decoded into the mel-spectrogram feature.



Note that the acoustic decoder can stop speech generation at the exact moment the synthetic speech reaches the length of the video clip, as the length of visual embedding indicates accurate utterance duration, thus avoid the infinite decoding problem in autoregressive speech synthesis.

\section{Experiments}
We conduct objective and subjective evaluation to assess the performance of VisualTTS for automatic voice over.  We note that there are no existing baselines for automatic voice over, so we propose to use two TTS baselines for comparison: Tacotron \cite{wang2017tacotron}, 
and a modified Tacotron with visual encoder and TVA, denoted as \textit{Tacotron with TVA}. We note that unlike VisualTTS, Tacotron with TVA has no visual fusion.
All baselines adopt the speaker encoder as described in Sec.~\ref{sec:spk} to support multi-speaker speech synthesis.



\subsection{Datasets and experimental setup}


We report the performance on GRID dataset \cite{cooke2006audio}, an audio-visual dataset consisting of 33 speakers, each speaking 1000 short English utterances. The training set consists of 900 sentences from 33 speakers, totaling 32,670 utterances. The remaining 100 sentences from each one of these 33 speakers are used for the test set. Speech audios are re-sampled at 24kHz and synchronized with 25Hz frame rate videos. 


We set the head number of the TVA to 2. TVA output is projected to 64 dimensions. The dimension of the visual fusion layer output is set to 256. The dimension of textual embedding is set to 512. The decoder RNN consists of 1 layer of attention RNN with 256-dimensional hidden size, and 2 layers of LSTM with 256-dimensional hidden size and 10\% zoneout rate. The acoustic decoder generates an 80-dimensional mel-spectrogram feature, two frames at a time, as output. The visual encoder is pre-trained on LRS2 and LRS3 datasets \cite{afouras2018deep, afouras2018lrs3}. The kernel size of Conv3D is $\{5,7,7\}$.
Visual embedding is a 512-dimensional vector for each frame of lip image sequence. Speaker embedding is a 256-dimensional d-vector obtained by a d-vector extractor pre-trained from a speaker verification task on AISHELL2 \cite{du2018aishell2} corpus. Speaker embedding is projected to 64 dimensions before concatenating with TVA output.
All models use WaveNet \cite{oord2016wavenet} pre-trained on VCTK dataset \cite{veaux2017cstr} as the vocoder for waveform generation.


\subsection{Objective evaluation}

We use Lip Sync Error - Confidence (LSE-C) and Lip Sync Error - Distance (LSE-D) \cite{prajwal2020lip} to measure lip-speech synchronization between silent videos from GRID dataset and synthetic speeches. 
We note that LSE-D measures the average distance between audio and lip representations obtained from a video of spoken utterance, while LSE-C is the average confidence score. LSE-C and LSE-D are measured using a pre-trained SyncNet \cite{Chung16a}. Lower LSE-D values and higher LSE-C values indicate better lip-speech synchronization.

LSE-C and LSE-D evaluation results are reported in Table \ref{tab:eval}. To start with, Tacotron with TVA and proposed VisualTTS both outperform Tacotron in terms of lip-speech synchronization. We note that VisualTTS achieves better synchronization than Tacotron with TVA. These results prove that both our visual-guided aligner and visual fusion strategy help to improve lip-speech synchronization. 





We use frame disturbance (FD) \cite{9262021} to measure duration deviation between synthetic speech and ground truth speech from the GRID dataset. We note that FD has been used to measure duration modeling performance of TTS~\cite{liu2021expressive}. Furthermore, as ground truth speech is synchronized with video, FD also indicates lip-speech synchronization between synthetic speech and video. VisualTTS achieves remarkable performance and outperforms both baselines with an FD value of 5.92.



\vspace{-5mm}
\subsection{Subjective evaluation}

We further conduct subjective evaluation to assess the performance of all three frameworks in terms of voice quality and lip-speech synchronization. 12 subjects participate in the listening tests, and each of them listens to 30 speech samples per framework.


We use mean opinion score (MOS) \cite{9262021} to appraise the voice quality. Each listener is asked to rate all speech samples on a five-point scale: higher scores indicate higher naturalness of speech samples. As shown in Table \ref{tab:eval}, all three frameworks achieve good voice quality and their performance are comparable to that of each other. 
We note that improving voice quality is not the main focus of VisualTTS. It is a TTS model that aims to achieve accurate  lip-speech  synchronization given text and video as input. 



\begin{figure}
    \centering
    \includegraphics[scale=0.42]{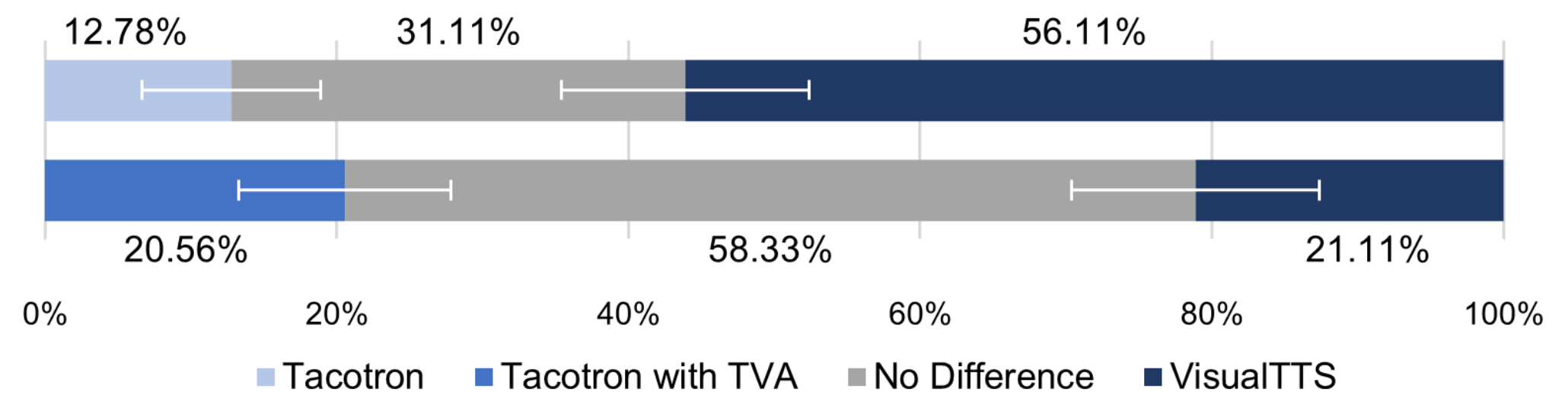}
    \vspace{-8mm}
    \caption{Preference test result for lip-speech synchronization with 95\% confidence intervals.}
    \label{fig:preference}
\end{figure}

We also conduct preference test on lip-speech synchronization. In this experiment, subjects are asked to watch each pair of videos and choose the one with better lip-speech synchronization. We note that we replace the original pre-recorded speeches in videos from the test set with synthetic speech samples produced by Tacotron, Tacotron with TVA, and VisualTTS. 
As shown in Fig.~\ref{fig:preference}, most of the subjects prefer videos with speech utterances synthesized by VisualTTS. These results prove the effectiveness of VisualTTS for generating speech samples that are in better synchronization with lip motion in videos. 


    

\begin{table}
    \caption{LSE-C, LSE-D, FD and MOS (with 95\% confidence intervals) evaluation results.}
    \centering
    \resizebox{0.49\textwidth}{0.13\linewidth}{
    \begin{tabular}{c|c c c c}
    \hline
     Method                      &  LSE-C $\uparrow$       &  LSE-D $\downarrow$ &  FD $\downarrow$  &  MOS $\uparrow$                      \\ \hline \hline
     Ground Truth                &  7.67        &  6.88  & NA   &  4.70$\pm$0.03            \\ \hline
     Tacotron \cite{wang2017tacotron}                   &  5.49        &  8.85   & 8.92  &  4.16$\pm$0.07            \\
     Tacotron with TVA             &  5.82        &  8.51   & 7.66  &  4.17$\pm$0.06             \\
     \textbf{VisualTTS}                    &  \textbf{5.87}        &  \textbf{8.45}  & \textbf{5.92}   &  \textbf{4.17$\pm$0.06}             \\
    
    \hline
    \end{tabular}}
    \label{tab:eval}
\end{table}

\vspace{-2mm}

\section{Conclusion}

\vspace{-2mm}



In this paper, we propose a new solution for AVO, introducing visual information to TTS for accurate lip-speech synchronization. We show that the proposed VisualTTS has a clear advantage over baselines in terms of lip-speech synchronization. As a future work, we are considering incorporating visual information with non-autoregressive TTS for more accurate lip-speech synchronization and fine-grained duration control with visual information.

{\footnotesize
\bibliographystyle{IEEEbib}

}
\end{document}